\documentstyle[12pt]{article}
\begin{document}

\begin{titlepage}
\begin{flushright}
{\tt Preprint YERPHI-1520(20)-98}\\
 {\tt hep-ph/9807411}
\end{flushright}

\vspace{4cm}

\begin{center}
  {\large\bf  Infrared Quasi-Fixed Points  and Mass Predictions in the
MSSM} \\[1cm]

  {\bf G. K.~Yeghiyan} \\[5mm]
  {\it  Yerevan Physics Institute,\\
       Yerevan,  Armenia} \\[1cm]

  {\bf  M. ~Jur\v{c}i\v{s}in\footnote{
On leave of absence from the Institute of Experimental Physics, SAS,
Ko\v{s}ice, Slovakia}  and  D. I.~Kazakov} \\[5mm]

{\it Bogoliubov Laboratory of Theoretical Physics,
Joint Institute for Nuclear Research, \\
141 980 Dubna, Moscow Region, Russian Federation}

\end{center}

\vspace{2cm}

\abstract{ We consider the infrared quasi-fixed point solutions of
the renormalization group equations for the  top-quark Yukawa
coupling and soft supersymmetry breaking parameters in the
MSSM. The IR quasi-fixed points together with the values of the
gauge couplings, the top-quark and  Z-boson masses allow one to
predict masses of the Higgs bosons, the stop squarks
and the lightest chargino as  functions of the only  free
parameter $m_{1/2}$ or the gluino mass.  The mass of the
lightest Higgs boson for $\mu>0$ and $M_{SUSY} \approx 1$ TeV is
found to be $m_h=(94.3+1.6+0.6\pm5\pm0.4)$ GeV. The case with
$\mu<0$ is excluded by experimental data.}

\end{titlepage}

\section{Introduction}

Supersymmetric extensions of the Standard Model  are believed to
be very promising theories to describe  physics at  energies to be
reached by experiments in the near future. The most popular SM
extension is the Minimal Supersymmetric  Standard Model (MSSM)
\cite{1}. Although the MSSM is the simplest supersymmetric model,
it contains a large number of free parameters.  The parameter
freedom of the MSSM comes mostly from soft supersymmetry breaking,
which is needed to obtain a phenomenologically acceptable mass
spectrum of particles. At the same time, a large number of free
parameters decrease the predictive power of a theory. A common way to
reduce this freedom is to make some assumptions at a high energy scale
(for example, at the Grand unification (GUT) scale or at the Planck scale).
Then, treating the MSSM parameters as running variables and using
the renormalization group equations (RGE's), one  can derive their
values at a low-energy scale.

The usual assumption is the so-called universality of
soft-breaking terms at high energy. Within a supergravity
induced SUSY breaking mechanism universality seems to be very natural
and leads at low energies to a softly broken  supersymmetric
theory which depends on the following set of free parameters \cite{1}:
a common scalar  mass $m_0$, a common gaugino mass  $m_{1/2}$, a common
trilinear scalar coupling $A$, a supersymmetric Higgs-mixing mass
parameter $\mu$, and a bilinear Higgs coupling $B$. These parameters
are defined at the high-energy scale and are treated as initial
conditions for the RGE's. The last two parameters can be eliminated in
favour of the electroweak symmetry breaking scale,
$v^2=v_{1}^2+v_{2}^2=(174.1\ GeV)^2$, and the Higgs fields vev's ratio
$\tan\beta=v_2/{v_1}$, when using minimization conditions of the Higgs
potential. The sign of $\mu$ is unknown and is a free parameter of the
theory.

Thus, using the concept of universality, one reduces the MSSM
parameter space to a five dimensional one. It is also possible to
restrict this remaining freedom. Namely, some low-energy MSSM
parameters are insensitive to the initial values.
This allows one to find their low-energy  values
without detailed knowledge of the physics at high energy.  To do this
one has  to examine  the infrared behaviour of RGE's for these
parameters and to find possible infrared fixed points.

Notice, however, that the true IR fixed points, discussed, e.g., in an
earlier
paper by Pendleton and Ross~\cite{2} are reached only in the asymptotic
regime. For the "running time" given by $\log M_{GUT}^2/M_Z^2$ they are
reached only by a very narrow range of solutions. This problem has
been resolved by consideration of more complicated fixed solutions
like invariant lines, surfaces, etc. \cite{3,4,5}. Such
solutions turned to be strongly attractive: for the
above-mentioned "running time" the wide range of solutions of RGE's ended
their evolution on these fixed manifolds.

Here we are interested in another possibility connected with
the so-called infrared quasi-fixed points first found out by  Hill \cite{6}
and then widely studied by several authors \cite{7}-\cite{12}.
These fixed points differ from  Pendleton-Ross ones at the intermediate
scale and usually give the upper (or lower) boundary for the  relevant
solutions.

Strictly speaking, these fixed points are not  always
relevant since the parameters of a theory do not necessarily have their
maximum or minimum allowable values.  For example, in the Standard
Model the Hill fixed point for the top-quark Yukawa coupling
corresponds to the pole top mass $m_t^{pole}=230$ GeV \cite{5},
which is excluded by modern experimental data~\cite{13}.

The situation is quite different in the MSSM~\cite{9}-\cite{12}:  here
the top-quark running mass is given by the equation
\begin{equation}
m_t=h_t v \sin{\beta}.
\end{equation}
Now the Hill fixed point  corresponds to physical values of the
top mass when $\tan{\beta}$ is chosen appropriately.

It is remarkable that imposing the constraint of bottom-tau
unification and radiative electroweak symmetry breaking leads to
the value of the top Yukawa coupling close to its quasi-fixed
point value~\cite{12,14}. This serves as an additional argument in
favour of the Hill-type quasi-fixed points. Therefore, in the
present paper we assume  $h_t(M_Z)$ to be equal to its Hill fixed
point  value. We restrict ourselves to the consideration of low
values of $\tan{\beta}$ for which one can neglect the bottom
Yukawa coupling as compared to the top one.

A similar analysis has been performed in a number of papers
\cite{11,15,16,17,18,19}. It has been pointed out that the IRQFP's
exist for the trilinear SUSY breaking parameter $A_t$  \cite{15},
for the squark masses \cite{11,16} and for the other soft
supersymmetry breaking parameters in the Higgs and squark sector
\cite{19}. This allows one to reduce the number of unknown parameters
and make predictions for the MSSM particle masses as
functions of the common scalar and gaugino masses $m_0$ and
$m_{1/2}$, respectively. In particular, in Ref.\cite{17} the
dependence of the lightest Higgs mass  on the ratio
$X=m_0/m_{1/2}$ has been studied.

In this paper, we perform a slightly different analysis. Based on
the values of IRQFP's for all soft terms and the known values of
gauge couplings, top-quark and Z-boson masses, we determine the
values of $\tan\beta$ and the $\mu$ parameter and then find masses
of the Higgs bosons, stops and charginos.  Our predictions are
insensitive (or weakly sensitive) to the initial values of
the ratios $m_0^2/m_{1/2}^2$ and $A_0/m_{1/2}$ in
the wide range of variations. Due to this insensitivity
the above-mentioned predictions depend only on one unknown
parameter, the gluino mass. We find also the mass of
the lightest Higgs boson as a function of the geometrical mean of
stops masses, sometimes called the SUSY breaking scale. These results
are confronted with the latest experimental data for searches of SUSY
particles and of the Higgs boson.

The paper is organized as follows. In Section 2 we analyze RGE's
and IR quasi-fixed points. We use them in Section 3 to obtain the
mass spectrum of the Higgs bosons and some SUSY particles. In
Section 4 we discuss our main results and conclusions. Appendix
contains some useful formulae.

\section{Infrared Quasi-Fixed Points  and RGE's}

In  this section, we give a short  description of the  infrared quasi-fixed
points (IRQFP) in the MSSM  for the low $\tan\beta$ scenario. In this case,
the only important Yukawa coupling is the top-quark one,
$Y_t=h_t^2/(4\pi)^2$.
It exhibits  the IRQFP behaviour  in the limit $Y_0=Y_t (0)
\to \infty$ ~\cite{6,9,10,11,15,17}
\begin{equation}
Y(t)\Rightarrow Y_{FP}=\frac{E(t)}{6 F(t)}, \label{fp} 
\end{equation}
where  $t=\log{M_{GUT}^2/Q^2}$ and the functions $E(t)$ and $F(t)$
are given in Appendix.

\begin{figure}[t]
     \vspace{-3.5cm}
       \begin{flushleft}
       \leavevmode
       \epsfxsize=7.5cm
       \epsffile{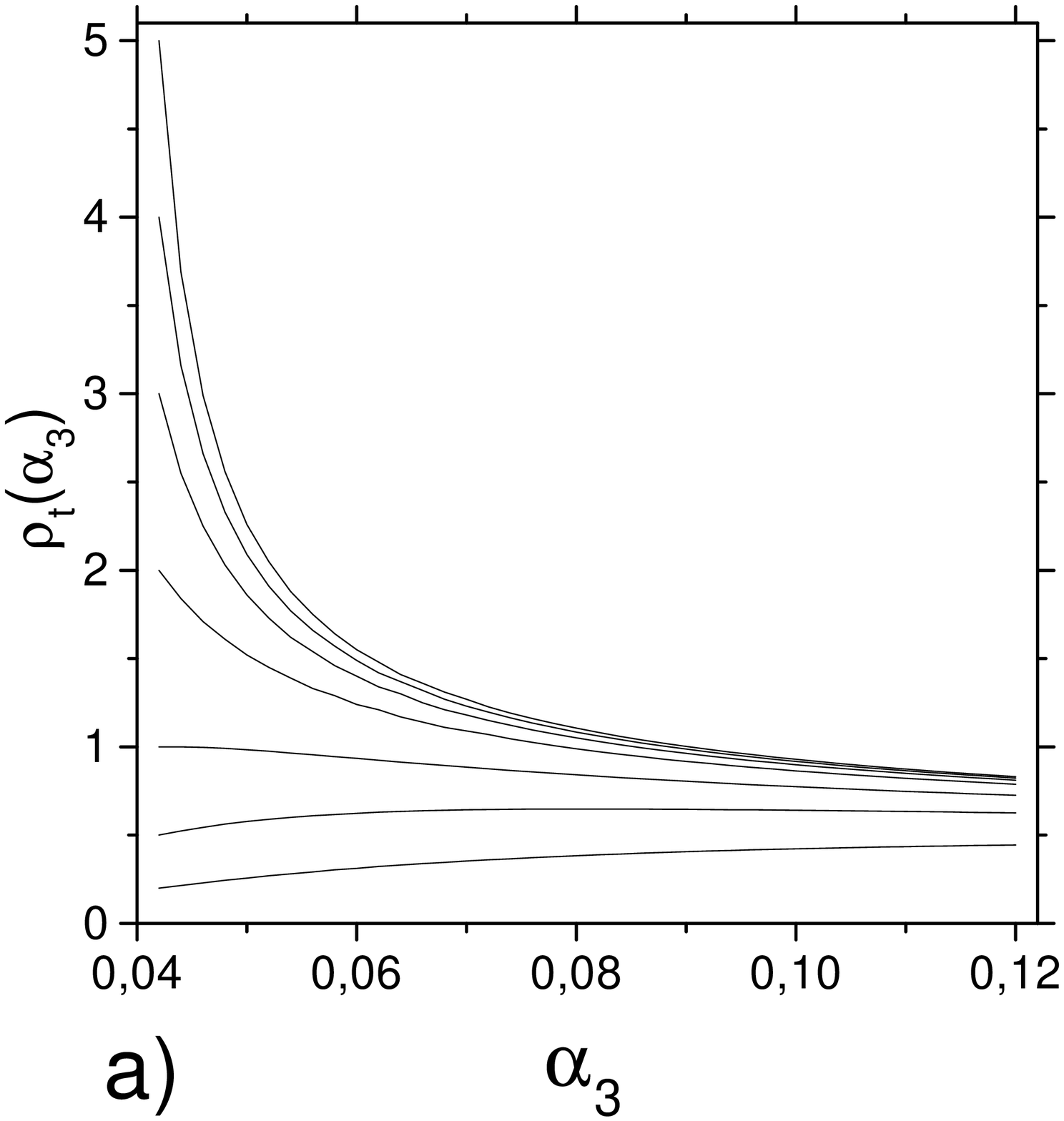}
   \end{flushleft}
     \vspace{-11.7cm}
   \begin{flushright}
       \leavevmode
       \epsfxsize=7.5cm
       \epsffile{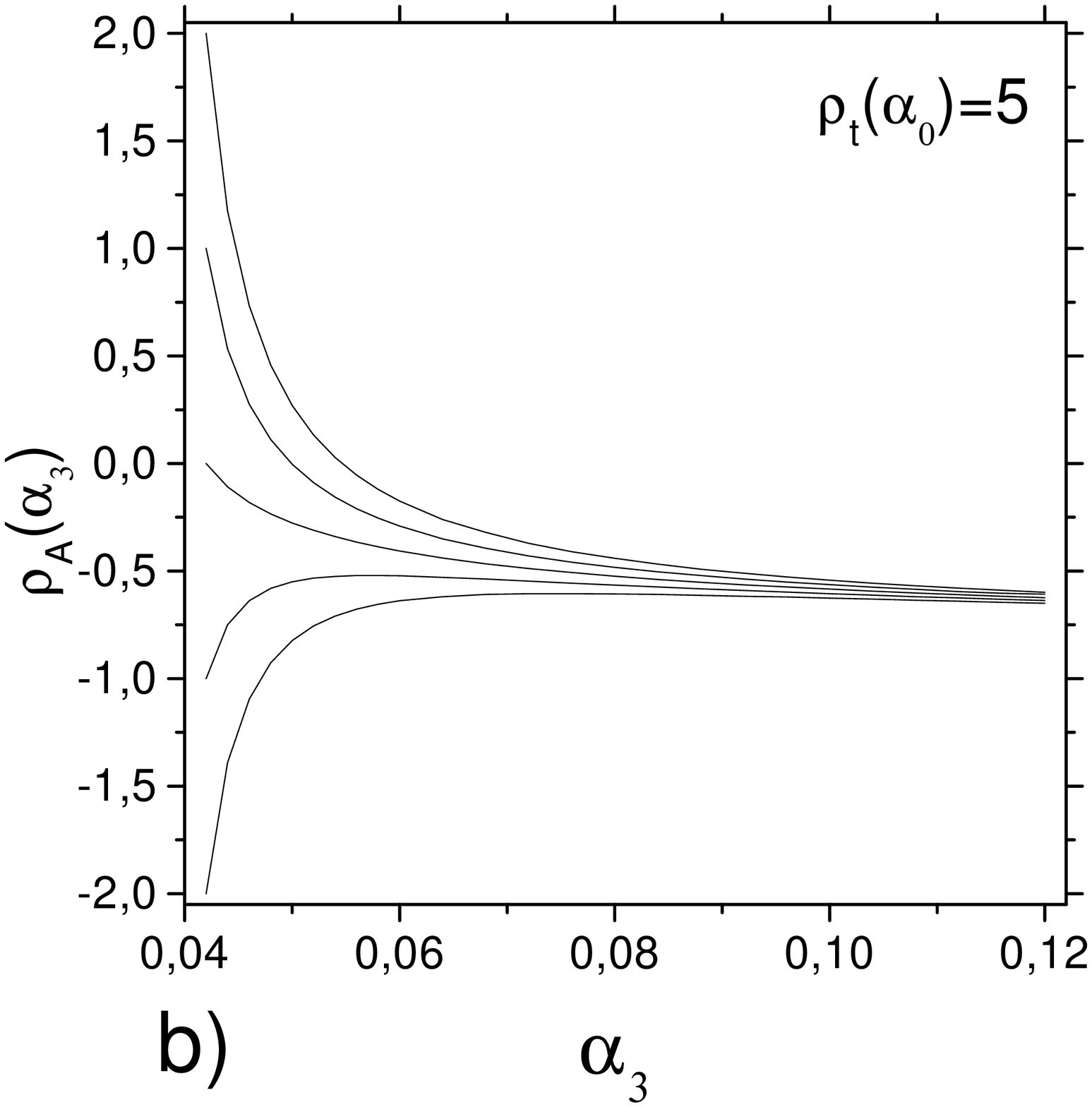}
   \end{flushright}
\caption{The infrared quasi-fixed points for
$\rho_t=Y_t/ \tilde\alpha_3 $ (a) and
$\rho_A=A_t/M_3$ (b)
\label{f1}}
\end{figure}

Though perturbation theory is not valid for $Y_t > 1$, it does not
prevent us from using the fixed point (\ref{fp}) since it attracts any
solution with $Y_0 \gg \tilde\alpha_{0}$ or, numerically, for $Y_0 >
0.1/4\pi$ (in fact, as one can see from Fig.\,\ref{f1}a, this
occurs for $Y_0 > 2 \tilde\alpha_{0}$). Here
$\tilde\alpha_i=\alpha_i/(4 \pi)$ and
$\tilde\alpha_0=\tilde\alpha_{i0}=\tilde\alpha_i(M_{GUT})$.
Thus, for a wide range of
initial values $Y_t$ is driven to the IR quasi-fixed point given by
eq. (\ref{fp}) which numerically corresponds to
$h_t(M_Z) \approx 1.125$. It is useful to introduce the ratio
$\rho_t \equiv Y_t/\tilde{\alpha}_3$. At the fixed point
$\rho_t(M_Z) \approx 0.84$.  The low-energy value of
$\rho_t$ is weakly dependent on  its high-energy value $\rho_0$
when the latter is big enough. This allows one to use  eq.(1)
to  predict  the top-quark mass as a function of $\tan{\beta}$
\cite{11}. Alternatively, one can use the IRQFP for $h_t$ to determine
$\tan{\beta}$ for a given value of the running top-quark mass.
We assume that the top  Yukawa coupling has its Hill fixed point value
$\rho_t(M_Z) \approx 0.84$ and analyze the IRQFP behaviour of the SUSY
breaking mass parameters in the MSSM with small $\tan{\beta}$.
It is worth  mentioning that our one-loop IRQFPs  differ
from those obtained in Ref.~\cite{19} due to ignorance
of the electro-weak corrections to the IRQFPs in this paper.
Contributions due to these corrections are more than 10\% and have to
be taken into account.

\input epsf

As it has been noted already, we are interested in predictions for
the Higgs bosons, the lightest chargino and the lightest squark (i.e.
stop) masses. These masses depend on the Higgs mixing parameter $\mu$, the
weak gaugino mass $M_2$, the soft masses from the Higgs
potential $m_{H_1}^2$ and $m_{H_2}^2$, the squark masses $m_Q^2$
and  $m_U^2$ (here Q refers to the third generation squarks
doublet and U to the stop singlet), the trilinear stop soft
coupling $A_t$ and $\tan{\beta}$. We determine $\tan{\beta}$ from
eq.(1).  The  parameter $\mu$ is eliminated from  the minimization
condition of the Higgs potential. As for the remaining parameters, we
express them via  the common gaugino mass $m_{1/2}$ or,
equivalently, via the gluino mass $M_3=(\tilde\alpha_3/
\tilde\alpha_{0}) m_{1/2}$, when  analyzing their IR fixed point
behaviour.

We perform the analysis of  RGE's at the one-loop level. The two-loop
corrections have negligible impact on our results. We neglect also
the effects connected with the mass thresholds and
consider the supersymmetric RGE's which are valid between the electroweak
breaking scale $\sim M_Z$ and the GUT scale $\sim 10^{16}$ GeV. Our
results are modified negligibly when the initial scale is chosen at
$M_{Planck} \sim 10^{19}$ GeV.

The relevant RGE's for soft supersymmetry breaking parameters
are given in Refs. \cite{22,14}.
The trilinear coupling $A_t$ in the limit $Y_0\to \infty$ possesses a
IR quasi-fixed point (see Appendix for the notation)
\begin{equation}
\frac{A_t(t)^{FP}}{M_3} = -\frac{\tilde\alpha_{0}}{\tilde\alpha_3}
\left(H_2-\frac{H_3}{ F(t)}\right) \approx -0.62\, \label{Atfp}
\end{equation}
independent of $A_t(0)=A_0$.

\input epsf
   \begin{figure}[t]
\vspace{-3.5cm}
  \begin{flushleft}
    \leavevmode
    \epsfxsize=7.5cm
    \epsffile{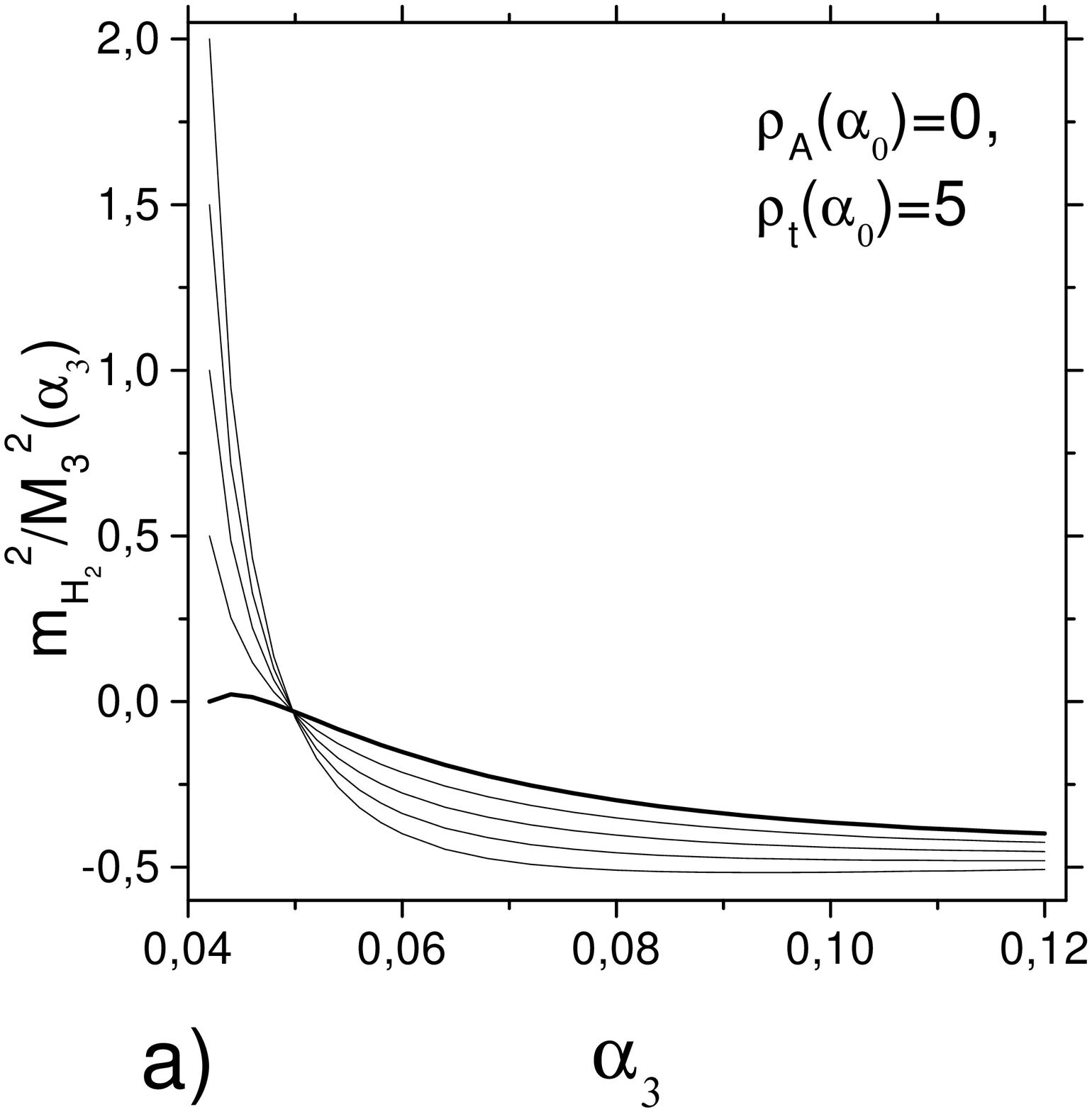}
  \end{flushleft}
\vspace{-11.7cm}
 \begin{flushright}
    \leavevmode
    \epsfxsize=7.5cm
    \epsffile{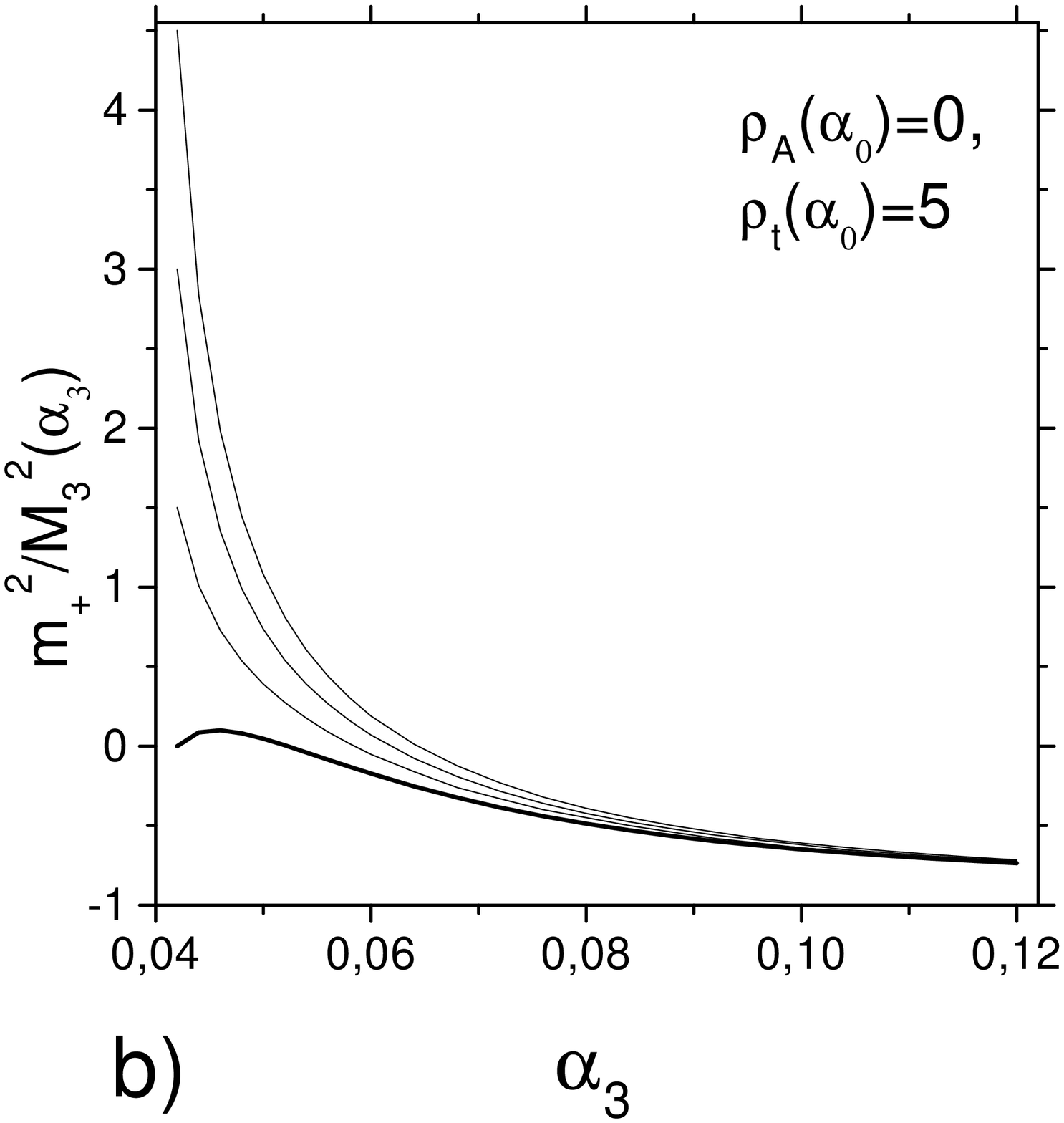}
\end{flushright}
\caption{The infrared quasi-fixed points for $m_{H_2}^2/M_3^2$ (a) and
    $(m_{H_1}^{2}+2 m_{H_2}^2)/M_3^2$ (b)
\label{f2}}
  \end{figure}
The behaviour of $\rho_A=A_t/M_3$ as a function of
$\tilde\alpha_3$ for  a fixed ratio $Y_0/\tilde\alpha_{0}=5$
is shown in Fig.\,\ref{f1}b.  One  can  observe the  strong attraction
to the IR stable quasi-fixed point $\rho_A \approx -0.62$. In fact, all
solutions corresponding to the small and moderate ratio $A_0/m_{1/2}$
"forget" their initial conditions.

The RGE's for $m_{H_1}^2$ and $m_{H_2}^2$ have the following solution
~\cite{22}:
\begin{eqnarray}
m_{H_1}^{2}&=&m_{0}^{2}+m_{1/2}^2 \tilde\alpha_{0}\left(\frac{3}{2} f_2(t)
+\frac{3}{10} f_1(t) \right)\,, \label{mh1}\\
m_{H_2}^{2}&=&m_{1/2}^{2} e(t) + A_t(0) m_{1/2} f(t)
+m_{0}^{2} h(t) - k(t) A_t(0)^2\,, \label{mh2}
\end{eqnarray}
where the functions $f_i(t)$, (i=1,2,3), $f(t)$, $h(t)$ and $k(t)$ are
given in Appendix.

There is no obvious infrared attractive fixed point
for $m_{H_1}^2$.  However, one can take the  linear combination
$m_+^2=m_{H_1}^2+2 m_{H_2}^2$  which together with $m_{H_2}^2$ shows
the IR fixed point behavior in the limit  $Y_0\to \infty$.
\begin{equation} \frac{m_+^{2\ FP}}{M_3^2}=
\left(\frac{m_{H_1}^2 + 2 m_{H_2}^2}{M_3^2}
\right)^{FP}=\frac{\tilde{\alpha}_{0}^2}{\tilde{\alpha}_3^2} \left(2 \bar
e(t) +
\frac 32 f_2(t)+\frac {3}{10} f_1(t)\right)
\approx -0.73\,. \label{h1}
\end{equation}
As for the ratio $m_{H_2}^2/M_3^2$, the solution for it has the
following form:
\begin{equation}
\frac{m_{H_2}^{2\
FP}}{M_3^2}= \frac{\tilde{\alpha}_{0}^2}{\tilde{\alpha}_3^2} \left(\bar
e(t) - \frac 12 \frac{m_{0}^{2}}{m_{1/2}^2}\right) \approx
-0.12 \left(\frac{1}{2}\frac{m_0^2}{m_{1/2}^2}+3.4 \right) \,.
\label{h2}
\end{equation}
In eq.(\ref{h2}) one has only  weak dependence on the ratio
$m_0^2/m_{1/2}^2$. This is because   $\bar e(t) \sim
\tilde\alpha_{3}^2/\tilde\alpha_0^2 \gg 1$. One can find the IR quasi-fixed
point $m_{H_2}^2/M_3^2 \approx -0.40$ which corresponds to
$m_0^2/m_{1/2}^2=0$.  As for the combination $m_+^2$, the dependence on
initial conditions disappears completely, as it follows from
(\ref{h1}). The situation is illustrated in Fig.\ref{f2}.

\input epsf
    \begin{figure}[t]
    \vspace{-3.5cm}
    \begin{flushleft} \leavevmode
    \epsfxsize=7.5cm
    \epsffile{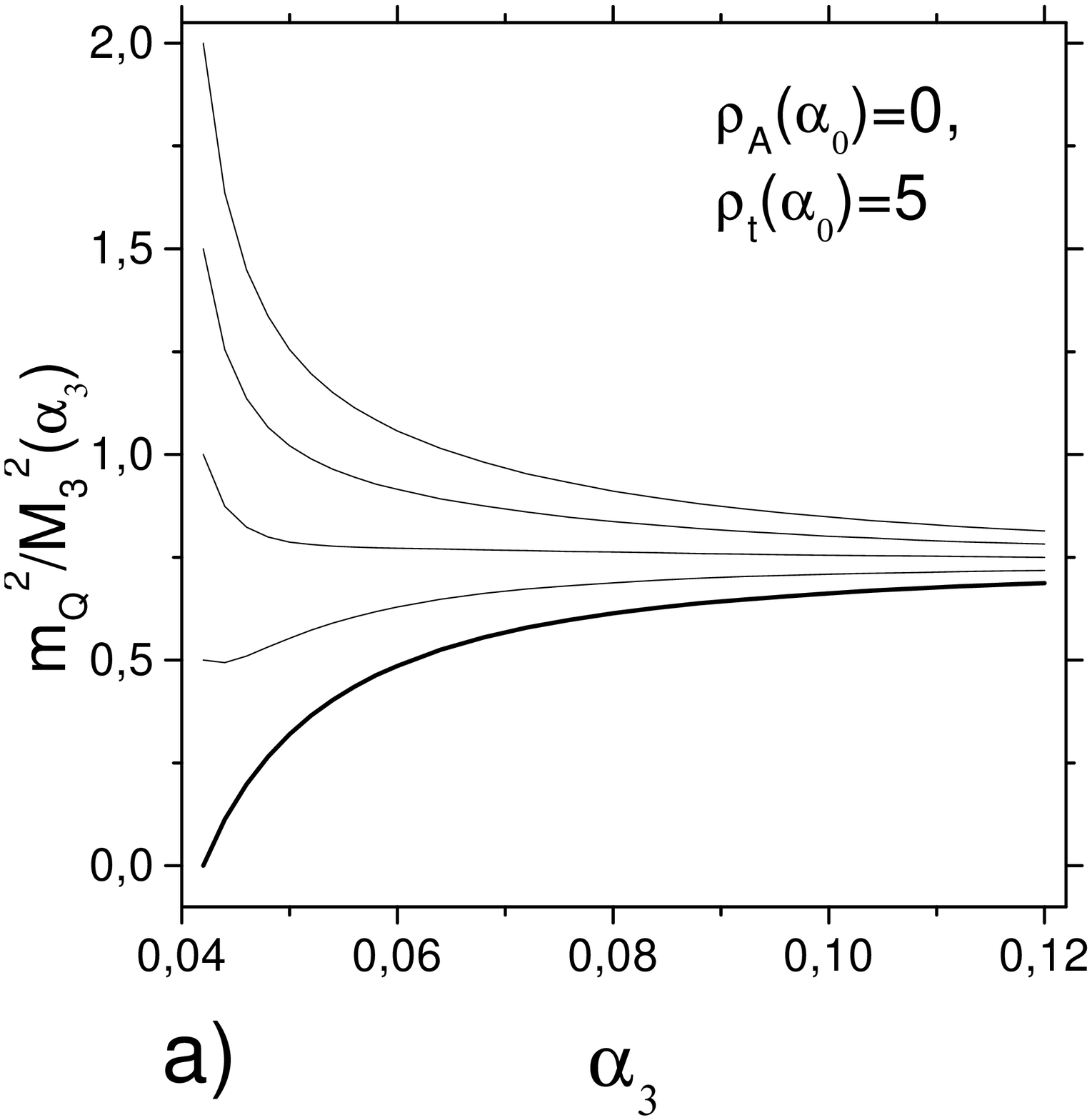}
  \end{flushleft}
\vspace{-11.7cm}
 \begin{flushright}
    \leavevmode
    \epsfxsize=7.5cm
    \epsffile{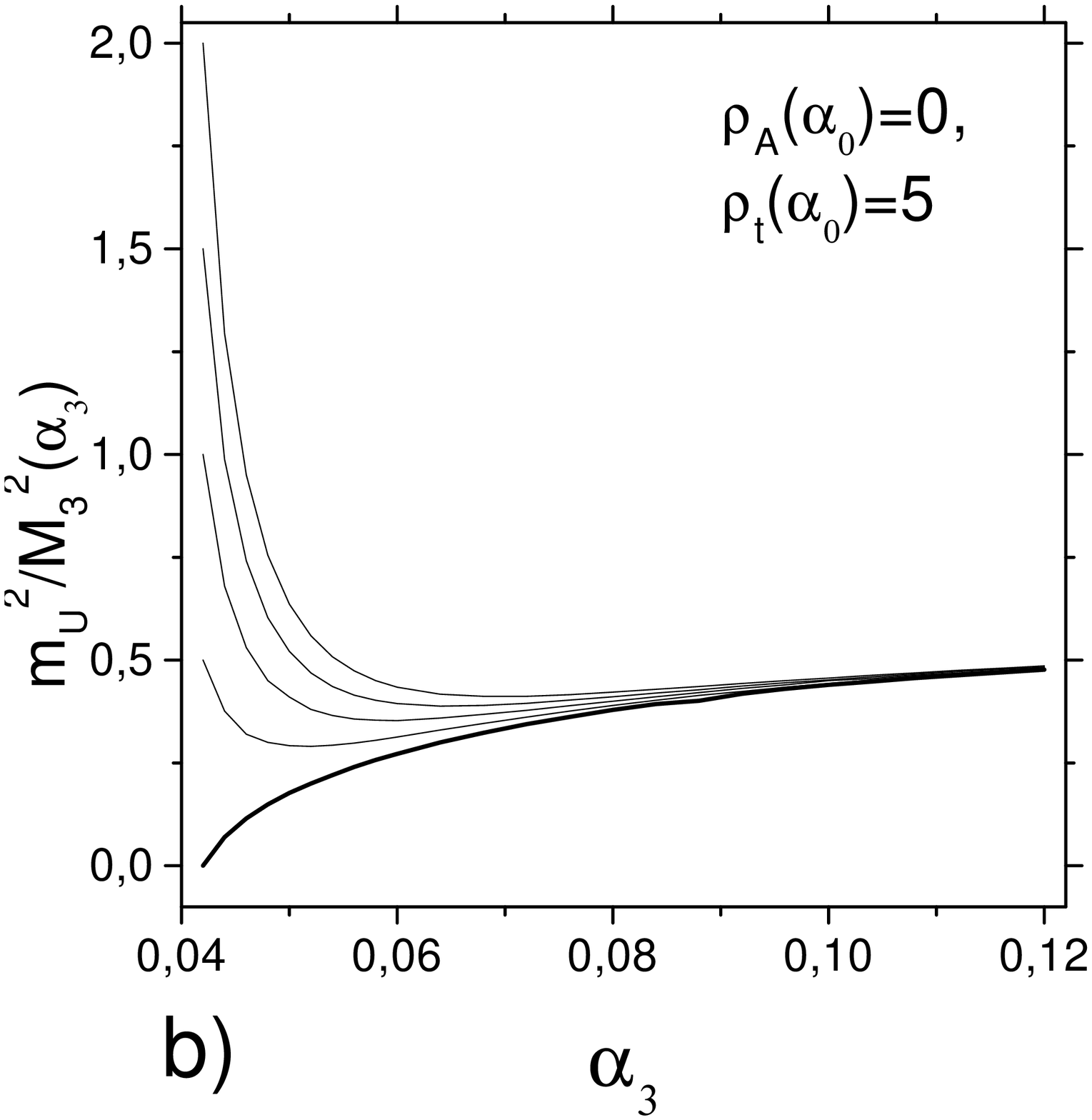}
\end{flushright}
\caption{The infrared quasi-fixed points for $m_{Q}^2/M_3^2$ (a) and
    $m_U^{2}/M_3^2$ (b)
\label{f3}}
  \end{figure}

Now we consider the squark masses. The solutions to their RGE's
are given in Appendix. In the limit $Y_t \to \infty$ these
solutions are driven to the IRQFP's
\begin{eqnarray}
\frac{m_U^{2\ FP}}{M_3^2} &=& \frac{\tilde{\alpha}_{0}^2}
{\tilde{\alpha}_3^2}
\left(\frac{2}{3} \bar{e}(t) + \frac{8}{3} f_3(t) - \frac{1}{2} f_2(t)
+\frac{13}{30} f_1(t) \right) \approx 0.48  \label{mufp} \\
\frac{m_Q^{2\ FP}}{M_3^2} &=& \frac{\tilde{\alpha}_{0}^2}{\tilde{\alpha}_3^2}
\left(\frac{1}{2} \frac{m_0^2}{m_{1/2}^2} +
\frac{1}{3} \bar{e}(t) + \frac{8}{3} f_3(t) + f2(t)
-\frac{1}{15} f_1(t) \right) \nonumber \\
&\approx& 0.12 \left(\frac{1}{2}
\frac{m_0^2}{m_{1/2}^2} + 5.8 \right) \label{mqfp}
\end{eqnarray}
As it follows from eq.(\ref{mufp}), the solution for
$m_{U}^2/M_3^2$ becomes independent of the initial conditions
$m_0/m_{1/2}$  and $A_0/m_{1/2}$, when the top-quark Yukawa
coupling is initially large enough. As a result, the solutions of
RGE's are driven to the fixed point $m_U^2/M_3^2 \approx 0.48$ for
a wide range of $m_0^2/m_{1/2}^2$ (Fig.\ref{f3}b). As for $m_Q^2$,
the dependence on initial conditions does not completely
disappear; however, it is rather weak like in the case of $m_{H_2}^2$ and
approaches the value $m_{Q}^2/M_3^2 \approx 0.69$
(Fig.\ref{f3}a).

Thus, one has two kinds of
infrared quasi-fixed points. For the first one ($m_+^2/M_3^2$
and $m_U^2/M_3^2$) the dependence
on initial values disappears completely when the
limit $Y_0 \to \infty$ is considered. As a
result, these fixed points attract any solution corresponding to small and
moderate values of the ratio $m_0^2/m_{1/2}^2$ (Fig. 2b, Fig. 3b).
For the second kind of IRQFP's
($m_{H_2}^2/M_3^2$ and $m_Q^2/M_3^2$)
weak dependence on initial values
persists; these fixed points exist due to the fact that
$\tilde{\alpha}_{3}^2/\tilde{\alpha}_{0}^2 \gg 1$. As a result,
one has small deviations of the RGEs solutions from these fixed points when
moderate values of the ratio $m_0^2/m_{1/2}^2$ are considered (Fig.
2a, Fig. 3a). In the next section we consider the
influence of such  deviations on the most
important predictions like the lightest Higgs boson
mass. We find out that
they have a
negligible impact on our results if small and moderate values of
$m_0^2/m_{1/2}^2 < 2$ are considered. This is due to
the fact that the lightest Higgs boson mass depends on supersymmetry
breaking parameters mainly via the stop squark masses which
are insensitive or weakly sensitive to the
above-mentioned deviations from the fixed points.

It is interesting to analyze also the behaviour of the bilinear
SUSY breaking parameter B. The determination of the ratio $B/M_3$
would allow one to eliminate the only remaining free parameter,
the gluino mass $M_3$, using the corresponding minimization
condition for the Higgs potential. However, the ratio $B/M_3$ does
not exhibit a fixed-point behaviour in the limit $Y_0 \gg
\tilde{\alpha}_{3}$. To see this, consider the
solution in the aforementioned limit. One has
\begin{equation} \frac{B}{M_3}
  \approx
0.35  \left(\frac{B_0}{m_{1/2}} - \frac{1}{2} \frac{A_0}{m_{1/2}}
-0.8 \right). \label{B}
\end{equation}
It is clear that neither $B_0/m_{1/2}$ nor $A_0/m_{1/2}$ may  be
neglected. As a consequence, no fixed point behaviour for the
ratio $B/M_3$  is observed.

Thus, the solutions of the RGE's for the soft
supersymmetry breaking  parameters are driven to the infrared
attractive fixed points when the top-quark Yukawa coupling
at the GUT scale is large enough.  Notice, however, that this is true if
only small and moderate values of the ratio $m_0^2/m_{1/2}^2$ are
considered. For larger values
of $m_0^2/m_{1/2}^2$ the above-mentioned fixed points
correspond
to the lower bounds for $m_+^2$, $-m_{H_2}^2=|m_{H_2}^2|$, $m_Q^2$
and $m_{U}^2$.  This can be easily understood, if one takes into
account that the above-mentioned fixed points are obtained in the formal
limit $m_0^2\to 0$ and $m_0^2$ is always positively defined.

In this paper, we restrict ourselves to the one-loop  RGE's. It is
interesting to see, however, how our results are modified when two-loop
RGE's are used. For comparison we present the two-loop IRQFP values
\cite{19} together with our one-loop results in Table 1.

\begin{table}[ht]
\begin{center}
\begin{tabular}{|c|c|c|c|c|c|}
\hline
 & $A_t/M_3$  & $m_U^2/M_3^2$ & $m_+^2/M_3^2$ & $m_{H_2}^2/M_3^2$ &
$m_Q^2/M_3^2$ \\ \hline
one-loop level & -0.62 & 0.48 & -0.73 & -0.40 & 0.69 \\ \hline
two-loop level & -0.59 & 0.48 & -0.72 & -0.46 & 0.75 \\ \hline
\end{tabular} \end{center}
\caption{ Comparison of one- and two-loop IRQFP's}
\end{table}

As one can see from this table, the difference between the
one-loop and two-loop  results is negligible for $A_t/M_3$,
$m_U^2/M_3^2$ and $m_+^2/M_3^2$. As for $m_{H_2}^2/M_3^2$ and
$m_Q^2/M_3^2$, the two-loop corrections to the fixed points are
about two times as small as deviations from them.
As it was mentioned above, such  corrections  have a negligible
impact on our main results.

We have discussed the case of low $\tan\beta$. For large
$\tan{\beta}$ the analysis becomes more complicated \cite{32}: one has to
take into account the parameters of a theory connected with the
b-quark and tau-lepton. One can follow the approach of Refs
\cite{3}-\cite{5} and consider more complicated fixed manifolds
like fixed lines, surfaces and multi-dimensional
subspaces.

\section{Masses of Stops, Higgs Bosons and Charginos}

In this section we use previously obtained  IR quasi-fixed points
for computation of masses of the Higgs bosons, stops and charginos.

First, we describe our strategy.  As input parameters we take the
known values  of the top-quark pole mass, $m_t^{pole}=(174.1\pm 5.4)$ GeV
\cite{13}, the experimental values of the gauge couplings \cite{20}
$\alpha_3=0.120 \pm 0.005$, $\alpha_2=0.034$, $\alpha_1=0.017$, the sum
of Higgs vev's squared $v^2=v_1^2+v_2^2=(174.1 GeV)^2$ and previously
derived fixed-point values for the top-quark Yukawa coupling and SUSY
breaking parameters. We use eq.(1) to determine $\tan{\beta}$ and the
minimization conditions for the Higgs potential to find the
parameter $\mu$. Then, we are left with a single free parameter,
namely $m_{1/2}$, which is directly related to the gluino mass
$M_3$. Varying this parameter within the experimentally allowed
range, we get all the masses  as functions of this parameter.

We start with determination of $\tan\beta$ which is related by
eq.(1) to the running top mass. The latter is found using its
well-known relation to the top quark pole mass including the
stop/gluino correction (see e.g. \cite{5,17} and references
therein)
\begin{equation}
m_t(m_t)=\frac{m_t^{pole}}{1+ \left(\frac{\Delta
m_t}{m_t}\right)_{QCD} + \left(\frac{\Delta m_t}{m_t}\right)_{SUSY}}.
\label{pole} \end{equation}
Eq.(\ref{pole}) includes the QCD gluon correction (in
the $\overline{MS}$ scheme)
\begin{equation} \left(\frac{\Delta
m_t}{m_t}\right)_{QCD}=\frac{4 \alpha_3}{3 \pi} + 10.92
\left(\frac{\alpha_3}{\pi}\right)^2\,,
\end{equation}
and the stop/gluino correction~\cite{21}
\begin{eqnarray} \nonumber
\left(\frac{\Delta m_t}{m_t}\right)_{SUSY} = &-&
\frac{g_3^2}{12 \pi^2} \Bigl\{B_1(m_t,M_3,\tilde m_{t_1}) +
B_1(m_t,M_3,\tilde m_{t_2})\\
&-& \sin(2\theta_t) \frac{M_3}{m_t} \left[B_0(m_t,M_3,\tilde m_{t_2})-
B_0(m_t,M_3,\tilde m_{t_1})\right] \Bigr\},
\end{eqnarray}
where $\theta_t$ is the stop mixing angle, $\tilde m_{t_1}<\tilde m_{t_2}$, and
\begin{eqnarray}
&&B_n(p,m_1,m_2)=-\int_0^1 dx\,x^n \ln\left[\frac{(1-x)m_1^2+x
m_2^2-x(1-x)p^2} {m_t^2} \right].
\end{eqnarray}
We use the following procedure to find the running top mass.
First, we take into account only the QCD correction and find
$m_t(m_t)$ at the first approximation.  This allows us to
determine both the stop masses and the stop mixing angle.
Next, having at hand the stop and gluino masses, we take
into account the stop/gluino correction. The result depends via
the stop masses on the sign of  $\mu$: one obtains $m_t(m_t) =
(162 \pm 5)$ GeV for $\mu >0$ and $m_t(m_t)=(165 \pm 5)$ GeV for
$\mu <0$.

Now we substitute the derived values of the top running mass into
eq.(1) to determine $\tan{\beta}$.  As a result, we obtain
$\tan{\beta}=1.47 \pm 0.15 \pm 0.05$ for $\mu>0$ and
$\tan{\beta}=1.56 \pm 0.15 \pm 0.05$ for $\mu<0$. The deviations
$\pm 0.15$ and $\pm 0.05$ from the central value are connected
with the experimental uncertainties of the top-quark mass and
$\alpha_3(M_Z)$, respectively. There is also some theoretical
uncertainty due to the fixed point value of $h_t(M_Z)$. It
has already been mentioned  that the solution of the RGE for the
top-quark Yukawa coupling is attracted by its Hill fixed point
when $\rho_0 > 2$. More precisely, one derives $h_t(M_Z)=1.09 \div
1.14$ when $2 < \rho_0 < 25$ (the upper bound on $\rho_0$
corresponds to $h_t^2(M_{GUT})=4\pi$, the perturbative limit).
Thus, for the wide range of initial values we obtain a very small
interval of $h_t(M_Z)$. In what follows, we take $h_t(M_Z) = 1.12$
derived for $\rho_0=5$. This value of $h_t(M_Z)$ has been used to
determine $\tan\beta$, as it was explained above. We must note,
however, that varying $h_t(M_Z)$ within the above-mentioned
interval one gains the uncertainty for $\tan\beta$ which is
comparable with the one coming from  the top mass, namely $\pm
0.1$ for both the cases $\mu>0$ and $\mu<0$. We take these
uncertainties into account when calculating the mass of the
lightest Higgs boson.

We would like to stress that due to the stop/gluino corrections to
the running top mass the predictions for $\tan\beta$ are different
for different signs of $\mu$. As a consequence the predictions for
the CP-odd and charged Higgs masses are also different for $\mu
>0$ and $\mu<0$ in spite of the fact that these parameters are not
explicitly dependent on the sign of $\mu$ at the tree level.

Further on we use only the central values for $\alpha_3=0.12$,
$m_t(m_t)=162$ GeV and $\tan{\beta}=1.47$ when $\mu>0$ and
$m_t(m_t)=165$ GeV, $\tan{\beta}=1.56$ when $\mu<0$ and  determine
the value of the parameter $\mu$. For this purpose, we use the
relation between  the Z-boson mass and the low-energy values of
$m^2_{H_1}$ and $m^2_{H_2}$, which comes from the minimization of
the Higgs potential
\begin{equation}
\frac{M_{Z}^{2}}{2}+\mu^2=\frac{m_{H_1}^{2}
-m_{H_2}^{2} \tan^2\beta}{\tan^2\beta-1}- \Delta_Z=
\frac{m_+^2
-m_{H_2}^{2}(\tan^2\beta+2)}{\tan^2\beta-1}- \Delta_Z\,, \label{mz}
\end{equation}
where $\Delta_Z$ stands for the one-loop corrections to the Higgs
potential and is given in Appendix (see ref.\cite{14} for details).
These corrections lower the value of $|\mu|$ by 10\% if $M_3>1$TeV.
This equation allows us to get the  absolute value of $\mu$,
the sign of  $\mu$ remains a free parameter.

Using eq.(\ref{mz}) and the IRQFP's for $m_+^2$ and $m_{H_2}^{2}$
evaluated in the previous section,
we are able to determine the mass
parameter $\mu^2$ as a function of the gluino mass $M_3$. At tree level
we obtain:  $\mu^2_{tree} \approx 0.80 M_3^2 - 4140.5$ for $\mu>0$ and
$\mu^2_{tree} \approx 0.74 M_3^2 -4140.5$ for $\mu<0$. As can be seen
from eq.(\ref{mz}) and previous discussion, fixed-point value
of $\mu^2$ corresponds to the lower bound
on this parameter as a function of $M_3$, when $Y_0 \gg
\tilde{\alpha}_{0}$. As for the gluino mass, it is restricted
only experimentally: for  arbitrary values of squark masses the gluino
mass is constrained by $M_3>173$ GeV \cite{26}.

Having all important parameters at hand, we are able now to
estimate the masses of phenomenologically interesting particles.

\input epsf
   \begin{figure}[t]
\vspace{-3.5cm}
  \begin{flushleft}
    \leavevmode
    \epsfxsize=7.5cm
    \epsffile{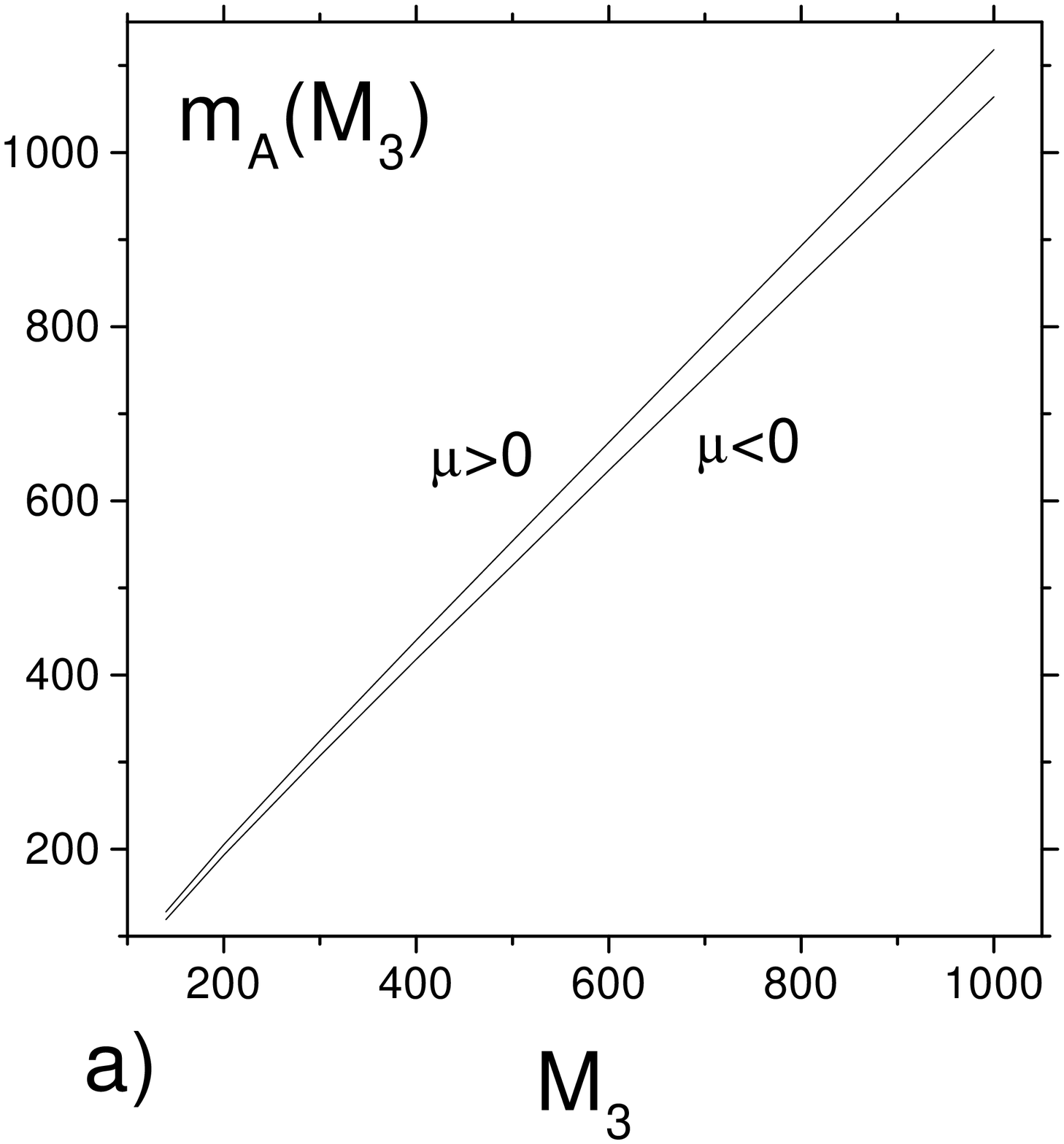}
  \end{flushleft}
\vspace{-10.7cm}
 \begin{flushright}
    \leavevmode
    \epsfxsize=7.5cm
    \epsffile{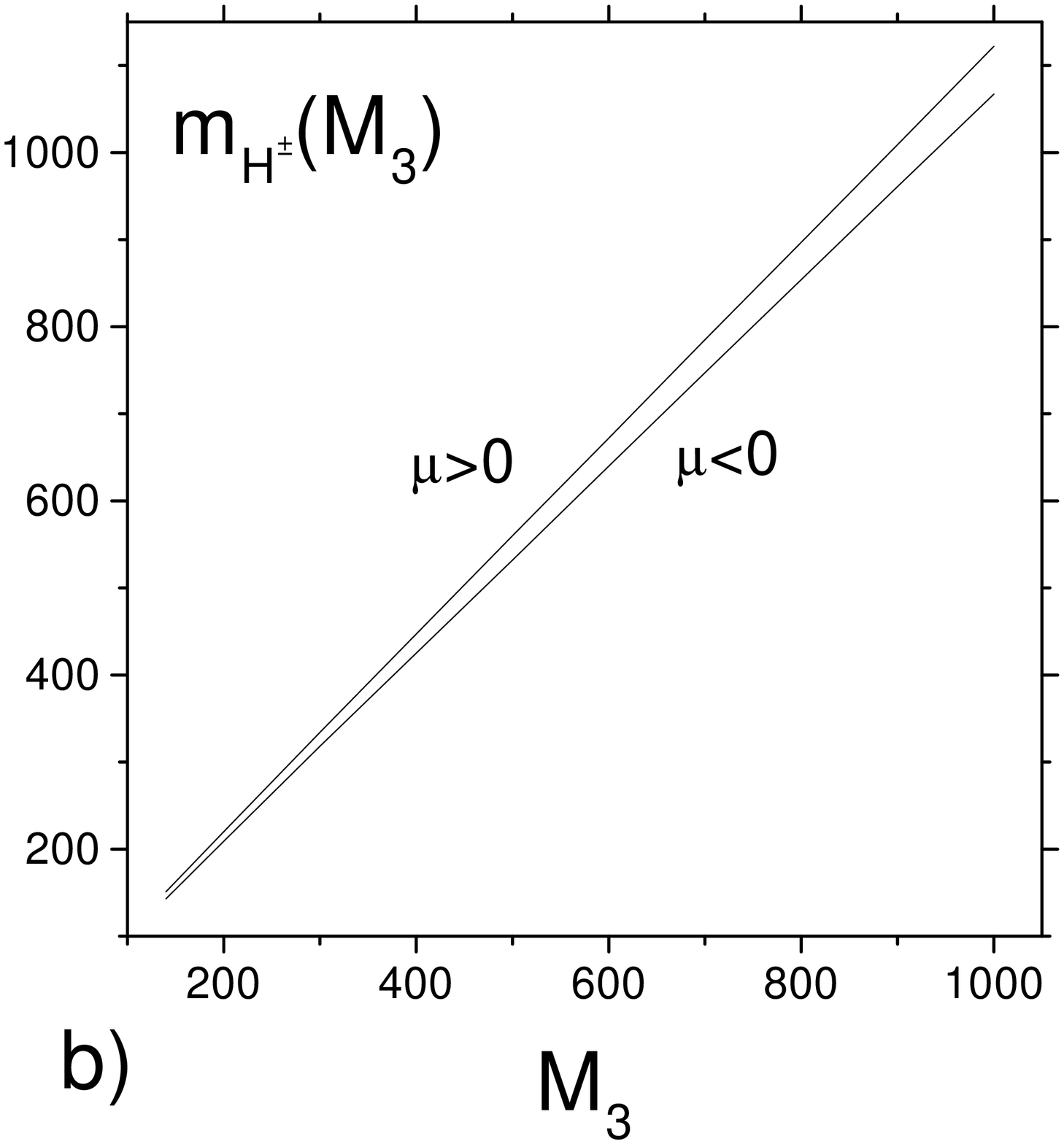}
\end{flushright}
\caption{ CP-odd (a) and charged (b) Higgs boson masses from the
IR quasi-fixed points for two signs of $\mu$ \label{f4}}
  \end{figure}

In the MSSM the Higgs sector consists of five physical states: two
neutral CP-even scalars $h$ and $H$, one neutral CP-odd scalar
$A$, and one complex charged Higgs scalar $H^{\pm}$. At the tree level
these particle masses are given by the following well-known
expressions \cite{10,14,27,28,31}:
\begin{eqnarray}
m_{A}^{2}&=&m_{H_1}^{2}+m_{H_2}^{2}+2 \mu^2_{tree}=m_+^2-m_{H_2}^2 + 2
\mu^2_{tree} \,, \label{a}
\\
m_{H^{\pm}}^{2}&=&m_{A}^{2}+M_{W}^{2}\,,
\\
m_{h,H}^2&=&\frac{1}{2} \left(M_Z^2 + m_A^2 \mp \sqrt{(M_Z^2+m_A^2)^2 -
4 M_Z^2 m_A^2 \cos^2{2 \beta}} \right)\,, \label{h}
\end{eqnarray}

For the CP-odd, charged and the heaviest CP-even Higgses we take the tree
level expressions because the radiative
corrections (including those to $\mu$) play a negligible role in this case
\cite{14,28,29,30}.
However, for the  lightest neutral Higgs boson the radiative
corrections are very important; therefore, we take them into account in
the two-loop order (see \cite{31}).  These corrections depend on stop
masses, trilinear soft coupling $A_t$ and $\mu$ calculated above.

\input epsf
   \begin{figure}[t]
\vspace{-3.5cm}
  \begin{flushleft}
    \leavevmode
    \epsfxsize=7.5cm
    \epsffile{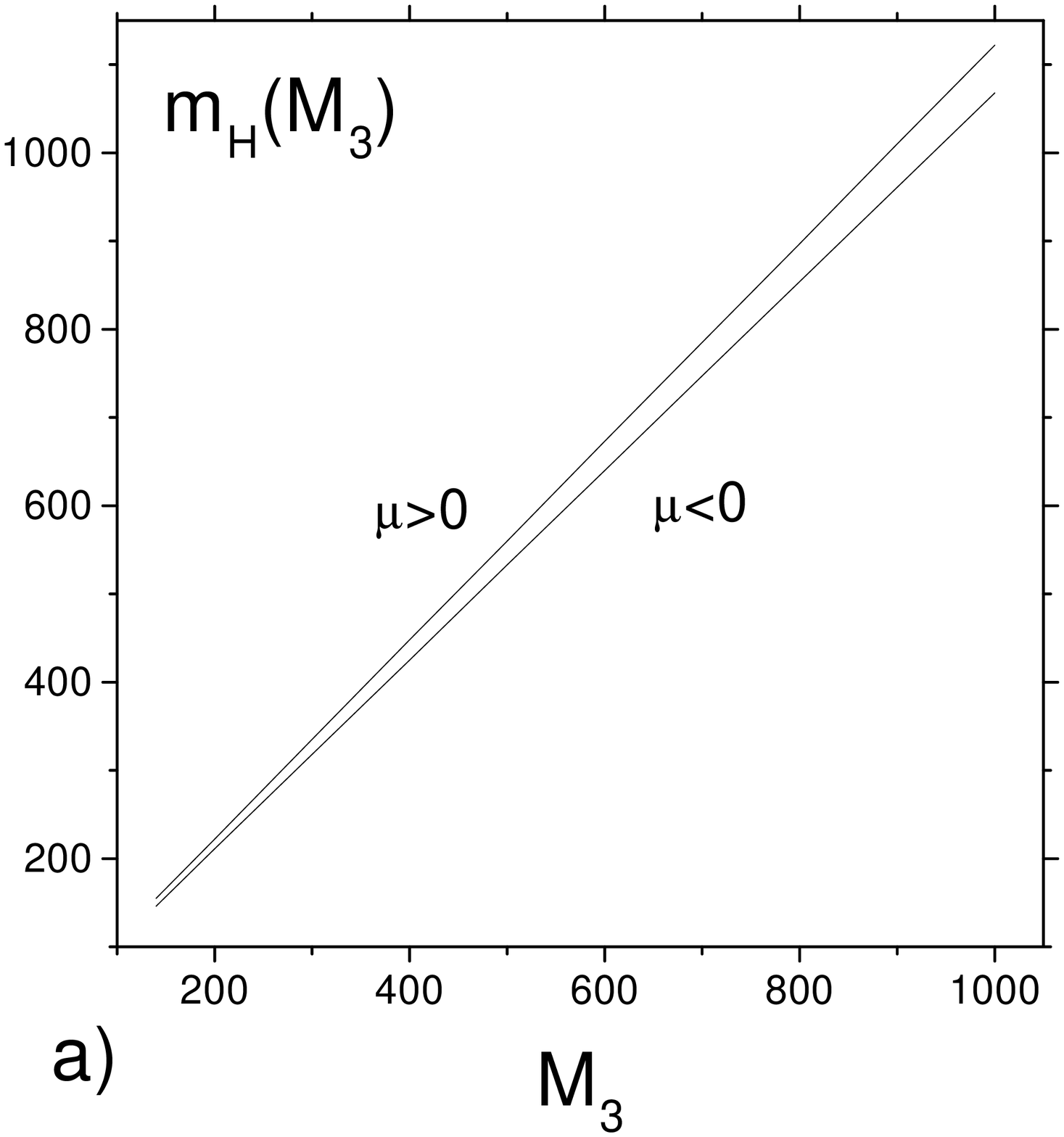}
  \end{flushleft}
\vspace{-11.7cm}
 \begin{flushright}
    \leavevmode
    \epsfxsize=7.5cm
    \epsffile{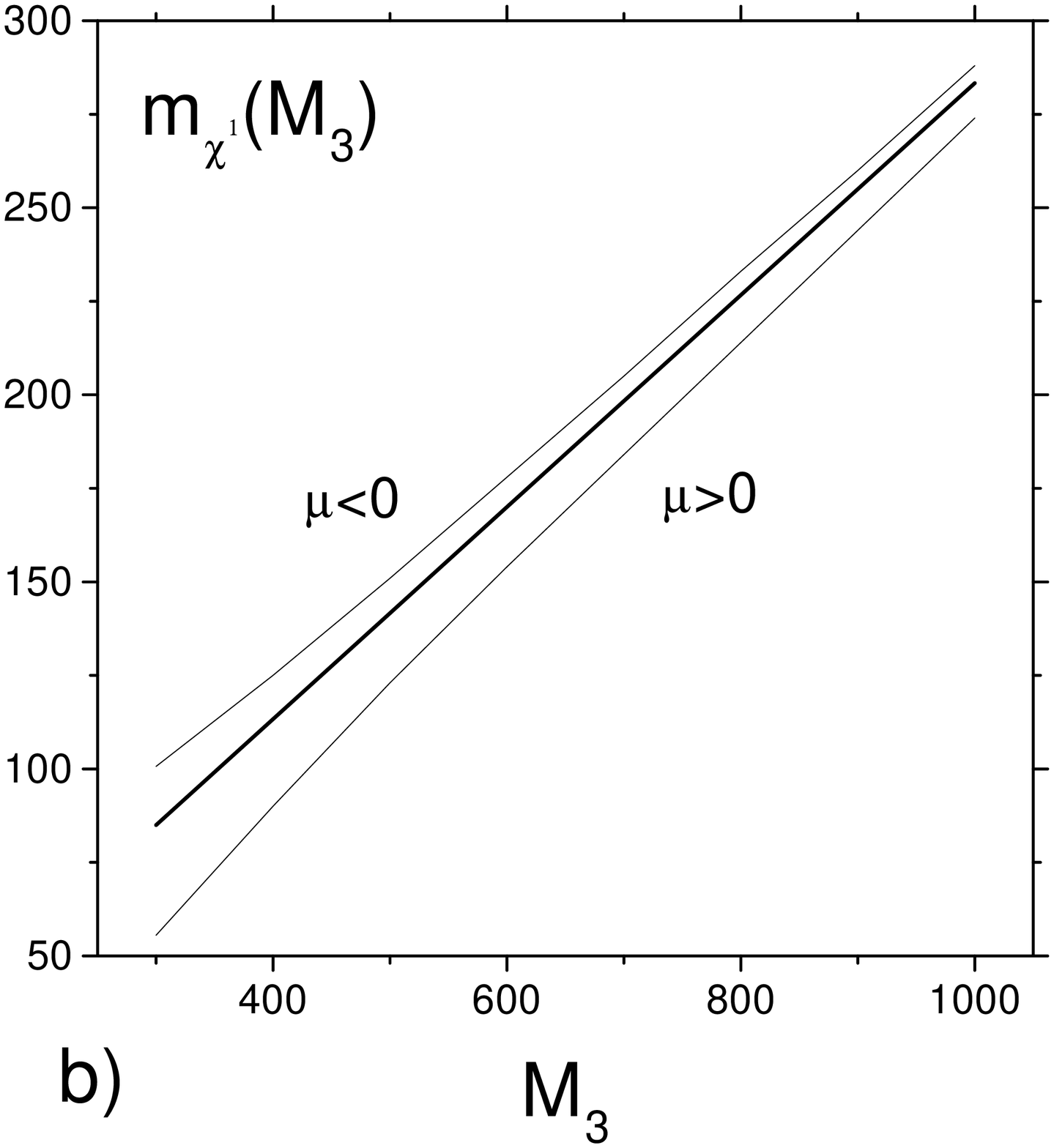}
\end{flushright}
\caption{ Heavy CP-even Higgs boson mass (a) and the chargino
mass (b) from the IR quasi fixed points for two signs of $\mu$.
Thick line corresponds to the limit $|\mu|\to\infty$ \label{f5}}
  \end{figure}

Using the values for $\mu^2$ and $\tan\beta$, we can calculate the
Higgs masses  as  functions of the gluino mass. We recall that the
fixed point values for $m_+^2$, $m_{H_2}^2$ and $\mu^2$ correspond
to the minimum values of these parameters in the limit $Y_0 \to \infty$.
This way the lower bounds for the
CP-odd, charged and heaviest CP-even Higgs boson masses as
functions of $M_3$ are obtained. These bounds are represented in Fig. 4
and Fig. 5a respectively.
As for the lightest
Higgs boson, we return to it after computation of the stop masses.

Now we proceed to the lightest chargino. The mass eigenvalues for
charginos  are the following~\cite{14}:
\begin{eqnarray}
m_{\chi_{1,2}}^{2}&=&\frac{1}{2} \big[ M_{2}^{2}+\mu^2 +2
M_{W}^{2} \nonumber
\\
&\mp& \sqrt{(M_{2}^{2}-\mu^2)^2+4 M_{W}^4\,\cos^{2}(2 \beta) +4
M_{W}^{2}(M_{2}^{2}+\mu^2+2 M_2 \mu \,\sin(2 \beta)}  \big] \,,
\label{ch}
\end{eqnarray}
where the weak gaugino mass is given by $M_2=(\tilde\alpha_2/
\tilde\alpha_3)M_3 \approx 0.23 M_3$.

We have all ingredients in eq.(\ref{ch}) to find the value of the
lightest chargino mass as  a function of $M_3$.  The result is
illustrated in Fig.\,\ref{f5}b for different signs of $\mu$. It
has already been mentioned that the IRQFP's for $m_+^2$ and
$m_{H_2}^2$ give the minimum value of $|\mu|$. We show also the
lightest chargino mass in the limit $|\mu| \to \infty$ (middle
line). Under the condition that $Y_0 \gg \tilde\alpha_{0}$, the lightest
chargino mass lies in the area between these three lines.
As one can see, the preferable value of the lightest chargino mass is
either its minimum or its maximum value, depending on the sign of
$\mu$.

\input epsf
   \begin{figure}[ht]
\vspace{-3.5cm}
  \begin{flushleft}
    \leavevmode
    \epsfxsize=7.5cm
    \epsffile{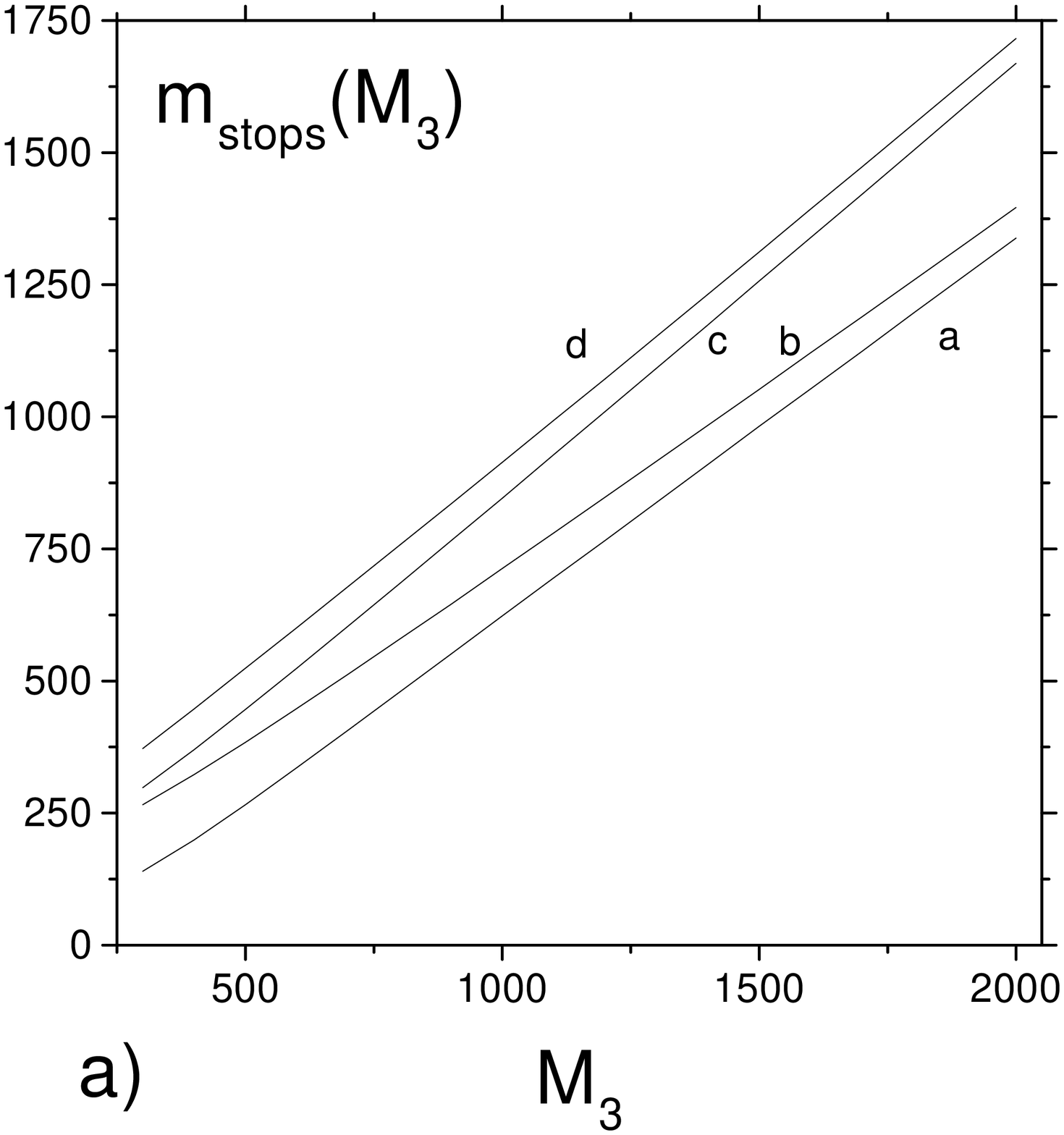}
  \end{flushleft}
\vspace{-11.7cm}
 \begin{flushright}
    \leavevmode
    \epsfxsize=7.5cm
    \epsffile{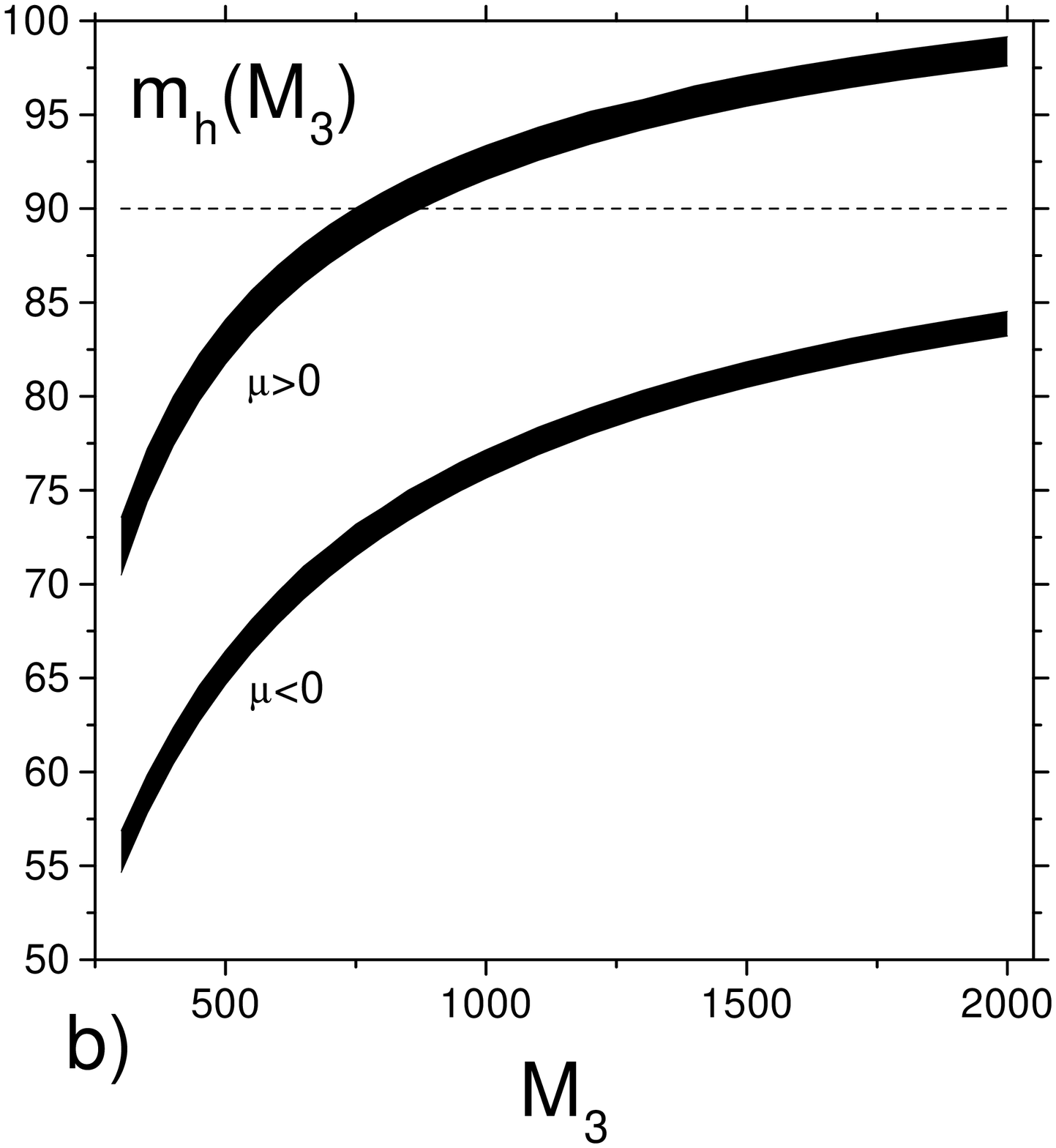}
\end{flushright}
\caption{ a) The values of the stop masses for two signs of $\mu$:
 $\tilde m_{t_1}$ for $\mu>0$ (a) and for $\mu<0$ (b),
 $\tilde m_{t_2}$ for $\mu<0$ (c) and for $\mu>0$ (d).
 b) The mass of the Higgs boson h for two signs of
$\mu$ (shaded areas). The dashed line correspond to a minimum value of
$m_h=90$ GeV, allowed by the experiment \label{f6}}
  \end{figure}

Consider at last the stop masses. After diagonalization of the
stop mass matrix its eigenvalues are given by the following
expression \cite{14}:
\begin{eqnarray}
\tilde m_{t_{1,2}}^{2}&=&\frac{1}{2} \big[\tilde m_{t_L}^{2}
+\tilde m_{t_R}^{2} \mp \sqrt{(\tilde m_{t_L}^{2} - \tilde
m_{t_R}^{2})^2 +4 m_{t}^{2} (A_t-\mu \, \cot \beta)^2} \big] \,,
\label{stop}
\end{eqnarray}
where $\tilde m_{t_L}^2$ and $\tilde m_{t_R}^2$ are given in
Appendix.

The behaviour of the stop masses as functions of the gluino mass is
presented in Fig.~\ref{f6}a. As one can see from this figure, both
$\tilde{m}_{t_1}$ and $\tilde{m}_{t_2}$
are of the order of the gluino mass (more precisely they vary
within the interval $0.6 M_3 - 1.5 M_3$).  We want to stress that such a
behaviour of the stop masses is the characteristic feature of the
IRQFP scenario. On the contrary, for $m_0 \gg m_{1/2}$
the soft parameters are far from the fixed points and
the lightest stop mass squared
may be arbitrary small or even negative \cite{32,8}. Such values of
$\tilde{m}_{t_1}^2$
are not possible when
fixed point values are taken.
Furthermore, comparing our results with those of Refs.
\cite{8,32}, it is easy to see that the fixed point value of the lightest
stop mass is very close to its maximum as a function of
$M_3$ when the
limit $Y_0 \gg \tilde{\alpha}_{0}$ is considered.
This explains why the stops are so heavy and have not been detected so far.
It suffices to note that one
obtains $\tilde{m}_{t_1} > 100$ GeV when using the experimental
bound on the gluino mass $M_3 > 173$ GeV. Further on we show
that for
the scenario at hand the lightest stop mass lies far out of the
domain of LEP~II when the experimental constraints on the
lightest Higgs mass are taken into account.

\input epsf
   \begin{figure}[ht]
\vspace{-3.5cm}
  \begin{center}
    \leavevmode
    \epsfxsize=7.5cm
    \epsffile{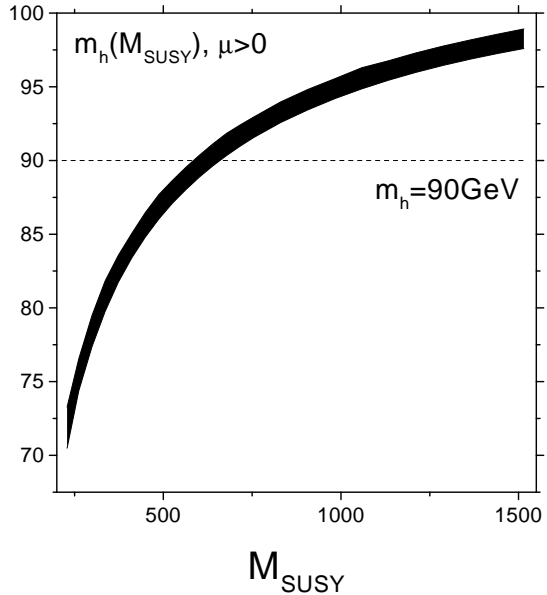}
  \end{center}
\caption{Dependence of the mass of the Higgs boson $h$ on
$M_{SUSY}=(\tilde m_{t_1} \tilde m_{t_2})^{1/2}$ (shaded area) for
$\mu>0$. The dashed line corresponds to a minimum value of
$m_h=90$ GeV, allowed by the experiment \label{f7}}
  \end{figure}

Now, when  the masses of stops are known , we can compute  the
mass of the lightest  Higgs boson. For this purpose we use
eq.(\ref{h}) together with one- and two-loop corrections \cite{31}. The
values of $\tilde m_{t_{1}}^{2}, \tilde m_{t_{2}}^{2}, A_t$ and
$\mu$ that enter the expressions for one- and two-loop corrections are
determined via IRQFP's for $m^2_Q,m^2_U, m^2_{H_2}$ and $m^2_+$. We
include also the deviations from the fixed points which are important
here. As it follows from Figs.\ref{f2} and \ref{f3} the fixed points for
$m^2_+$ and $m^2_U$ have a very strong attraction and are reached
for a wide range of initial conditions. For this reason we take
the fixed point values for these masses. On the contrary,  the
fixed points for $m^2_{H_2}$ and $m_Q^2$ are less attractive and
we take into acccount the deviations from them. In our numerical
analysis we took $m^2_{H_2}/M_3^2=-0.52\div -0.40$ and
$m_Q^2/M_3^2=0.69\div 0.81$. These intervals determine the
uncertainty of our evaluation of the Higgs mass. The values
$m^2_{H_2}/M_3^2=-0.40$ and $m_Q^2/M_3^2=0.69$ correspond to
$m_0^2/m_{1/2}^2=0$, and the values $m^2_{H_2}/M_3^2=-0.48$ and
$m_Q^2/M_3^2=0.81$ correspond to $m_0^2/m_{1/2}^2=2$.

Fig.\,\ref{f6}b shows the dependence of $m_h$ on the gluino
mass for two signs of $\mu$. The shaded areas correspond to the
deviations from the fixed points discusses above.  As one can see
from this picture, the IRQFP scenario excludes the negative sign of
$\mu$ in agreement with the conclusions made in Ref.~\cite{37}
(see also discussion in the end of this section).

Further on we consider only the positive values of $\mu$.
Fig.\,\ref{f7} shows the value of $m_h$ for $\mu>0$ as a function
of the geometrical mean of stop masses - this parameter is often
identified  with  a supersymmetry breaking scale $M_{SUSY}$.
For $M_{SUSY}$ of the order of 1 TeV the value of the lightest
Higgs mass is $m_h=(94.3+1.6+0.6)$ GeV, where the first uncertainty  comes
from the deviations from the fixed points for the mass
parameters and the second one is related to that of the
top-quark Yukawa coupling.
This value of $m_h$ is calculated for the
top-quark mass equal to 174.1 GeV. If one adds the uncertainty of
the determination of the top-quark mass of 5.4 GeV, one finds an
almost linear dependence of $m_h$ on $m_t^{pole}$. The last source
of uncertainties is the value of $\alpha_3(M_Z)$. If one takes
the value  $\alpha_3(M_Z)=0.120 \pm 0.005$, it gives the
uncertainty in the Higgs mass of $\pm 0.4$ GeV.
Finally, one has the following prediction for the lightest
Higgs boson mass:
\begin{equation}
m_h=(94.3+1.6+0.6\pm5\pm0.4) \ \mbox{GeV}, \ \ \ \mbox{ for} \ M_{SUSY}=1
\ TeV. \label{mass}
\end{equation}
One can see that the main source of
uncertainty is the experimental error in the top-quark mass. As
for the uncertainties connected with the fixed points, they give
much smaller errors of the order of 1 GeV.

Note  that the  obtained result (\ref{mass}) is very close to the
upper boundary, $m_h=97$ GeV, reported in Refs. \cite{17,44}. This means
that using the IRQFP approach for all
relevant SUSY
breaking mass parameters, one can explain in the natural way, why the
lightest Higgs boson was not detected so far. We recall that the similar
result was obtained also for the lightest stop mass.

Thus, for a supersymmetry breaking scale of an order of 1 TeV the
IRQFP scenario gives the lightest Higgs boson mass within the
reach of LEPII. The absence of the Higgs boson events up to 100
GeV would indicate either larger values of the SUSY breaking scale
or higher values of $\tan\beta$. For $M_{SUSY} > 1$ TeV one
has to perform a more detailed analysis of radiative corrections
taking into account the running of the top mass and QCD coupling
constant between the electroweak scale and SUSY scale. For
higher values of $\tan\beta$ the analysis of fixed solutions
becomes more complicated, due to new Yukawa couplings being
involved.  One may consider, of course, more radical changes
like non-universality of the soft terms, new particles
which appear for instance in gauge-mediated SUSY breaking models,
etc.

The experimental bounds play an important role for the remaining
particles as well.
It has been noted already  that the experimental bound on gluino mass
pushes the lightest stop to be heavier than 100 GeV.
The restrictions on particles masses become stronger
when the experimental constraint on the lightest Higgs boson is
taken into account. In the MSSM besides the Standard Model
channel $e^+ e^- \rightarrow h Z$ there is an additional channel
for the Higgs boson production, $e^+e^- \rightarrow h A$ (see e.g.
\cite{34,35}). The modern experimental constraint on the Higgs mass in
the MSSM is $m_h > 62.5$~GeV \cite{PDG}.  We have to recall, however,
that the CP-odd mass has been obtained here to be larger than the
Z-boson mass. This means that the Standard Model constraint on the
Higgs mass $m_h > 90$ GeV \cite{36} can be used also for the minimal
supersymmetry.

Let us see how this constraint affects the values of
sparticles masses. First,  one can see from  Fig.\ref{f7} that
$\sqrt{\tilde m_{t_1} \tilde m_{t_2}}$ must be larger than 550 GeV
in order the condition $m_h > 90$ GeV to be satisfied. For the IRQFP
scenario considered here this is equivalent to restriction on
the gluino mass $M_3 \geq 750$ GeV (see Figs.\,\ref{f6},\ref{f7}).
Subsequently one obtains $m_A >844$ GeV, $m_{H^\pm}>846$ GeV, $m_H
> 848$ GeV, $\tilde{m}_{t_1} > 440$ GeV for $\mu>0$.
As for the lightest chargino, one has $m_{\chi_1} \geq 200$ GeV.
Thus, these particles are too heavy to be detected in the nearest
experiments.

\section{Summary and Conclusion}

We have analyzed the fixed point behaviour of the SUSY breaking
parameters in the small $\tan{\beta}$ regime. We have made this
analysis, assuming that the top quark Yukawa coupling is initially large
enough to be driven at the infrared scales to its Hill-type quasi-fixed
point or, equivalently, to its upper boundary value. This value of
$h_t(M_Z)$ corresponds to a possible Grand Unification scenario with
bottom-tau unification and radiative EWSB.

We have found  that solutions of RGE's for some of the SUSY breaking
parameters become insensitive to their initial values at unification
scales. This is because at the infrared scales they are driven to their
IR quasi-fixed points. These fixed points are used to
make predictions for the masses of the Higgs bosons, stops and the
lightest chargino as the functions of the only unknown parameter -
the gluino mass.   We have taken into account possible impact of
deviations from  the IR quasi-fixed points as well as
experimental bounds on the gluino and the lightest Higgs masses.

For the  infrared quasi-fixed point scenario the Higgs bosons
except the lightest one are found to be too heavy to be
accessible in the nearest experiments. The same is true for the
stops and charginos.
This explains in a natural way, even without knowlegde of physics
at unification scale, why the stops and charginos have not been
detected so far. The mass of
the lightest neutral Higgs boson
is also close to its upper boundary, however,
for the low $\tan\beta$ case it is within the reach of LEP~II.

\section*{Appendix }
\underline{\large{1. Notation}} \cite{22}
\begin{eqnarray*}
&& b_1=11,\ b_2=1,\ b_3=-3,\  t=log\frac{M_{GUT}^2}{Q^2} \\
&& \beta_i=b_i \tilde{\alpha}_{i0},\ i=1,2,3 \\
&&E(t)=(1+\beta_3 t)^{16/(3b_3)}(1+\beta_2 t)^{3/b_2}
(1+\beta_1 t)^{13/(15b_1)} \,, \\
&&F(t)=\int\limits_0^t E(t') dt' \,, \\
&& D(t)=1+6 Y_0 F(t) \,, \\
&& f_i(t)=\frac{1}{\beta_i}\left( 1-\frac{1}{(1+\beta_i t)^2} \right)\,, \\
&&h_i(t)=\frac{t}{(1+\beta_i t)}\,, \\
&&e(t)=\frac{3}{2}\left[
\frac{G_1(t)+Y_0 G_2(t)}{D(t)}+
\frac{(H_2(t)+6 Y_0 H_4(t))^2}{3 D^2(t)}+H_8(t) \right]\,, \\
&& \bar e(t)=\frac{G_2(t)}{4 F(t)} +
\frac{H_4^2(t)}{2 F^2(t)} +
\frac{3 H_8(t)}{2} \,, \\
&&f(t)=-\frac{6 Y_0 H_3(t)}{D^2(t)}\,, \\
&&h(t)=\frac{1}{2}\left(\frac{3}{D(t)}-1 \right)\,, \\
&&k(t)=\frac{3 Y_0 F(t)}{D^2(t)}\,, \\
&&H_2(t)=\left(\frac{16}{3} \tilde\alpha_{30} h_3(t) + 3 \tilde\alpha_{20}
h_2(t) + \frac{13}{15} \tilde\alpha_{10} h_1(t) \right)\,, \\
&&H_3(t)=t E(t)-F(t)\,, \\
&&H_4(t)=F(t) H_2(t)-H_3(t)\,, \\
&&H_5(t)=\left(-\frac{16}{3} \tilde\alpha_{30} f_3(t) + \tilde\alpha_{20} 6
f_2(t) -\frac{22}{15} \tilde\alpha_{10} f_1(t) \right)\,, \\
&&H_6(t)=\int\limits_0^tH_{2}^2(t')E(t')dt'\,, \\
&&H_7(t)=\left(3 \tilde\alpha_{20} h_2(t)
+\frac{3}{5} \tilde\alpha_{10} h_1(t) \right)\,, \\
&&H_8(t)=\left(-\frac{8}{3} \tilde\alpha_{30} f_3(t) +
\tilde\alpha_{20} f_2(t) -\frac{1}{3} \tilde\alpha_{10} f_1(t) \right)\,, \\
&&G_1(t)=F_2(t)-\frac{1}{3}H_{2}^2(t)\,, \\
&&G_2(t)=6 F_3(t)-F_4(t)-4 H_2(t) H_4(t)
+2 F(t) H_{2}^2(t)-2 H_6(t)\,, \\
&&F_2(t)=\left(\frac{8}{3} \tilde\alpha_{30} f_3(t)
+ \frac{8}{15} \tilde\alpha_{10} f_1(t) \right)\,, \\
&&F_3(t)=F(t)F_2(t)-\int\limits_0^tE(t')F_2(t')dt'\,, \\
&&F_4(t)=\int\limits_0^tE(t')H_5(t')dt'\,,
\end{eqnarray*}
\vspace{0.2cm} \\
\underline{\large{2. Solutions of the RGE's for supersymmetry breaking
parameters \cite{14,22}}}
\begin{displaymath}
A_t(t) = \frac{A_t(0)}{1+6 Y_0 F(t)}
-m_{1/2}\left(H_2-\frac{6 Y_0 H_3}{1+6 Y_0 F(t)}\right)\, , \label{At}
\end{displaymath}
\begin{eqnarray*}
m_{H_1}^{2}&=&m_{0}^{2}+m_{1/2}^2 \left(\frac{3}{2} \tilde\alpha_{20}
f_2(t) + \frac{3}{10} \tilde\alpha_{10} f_1(t) \right)\,, \\
m_{H_2}^{2}&=&m_{1/2}^{2} e(t) + A_t(0) m_{1/2} f(t)
+m_{0}^{2} h(t) - k(t) A_t(0)^2\,,
\end{eqnarray*}
\begin{eqnarray*}
m_Q^2(t)=\frac{2}{3} m_0^2+\frac{1}{3} m_{H_2}^2+m_{1/2}^2
(\frac{8}{3} \tilde\alpha_{30} f_3(t) + \tilde\alpha_{30} f_2(t) -
\frac{1}{15} \tilde\alpha_{10} f_1(t)) \\
m_U^2(t)=\frac{1}{3} m_0^2+\frac{2}{3} m_{H_2}^2+m_{1/2}^2
(\frac{8}{3} \tilde\alpha_{30} f_3(t) - \tilde\alpha_{20} \frac{1}{2} f_2(t)
- \frac{13}{30} \tilde\alpha_{10} f_1(t))
\end{eqnarray*}
The parameters $\tilde m_{t_L}^2$ and $\tilde m_{t_R}^2$ are connected
with $m_{Q}^2$ and $m_{U}^2$ by the following relations:
\begin{eqnarray*}
\tilde m_{t_R}^{2}&=&m_{U}^{2} +
m_t^2 + M_Z^2 \cos{2\beta} (\frac{2}{3}
\sin^2{\theta_W}) \\
\tilde m_{t_L}^{2}&=&m_{Q}^{2} +
m_t^2 - M_Z^2 \cos{2\beta} (\frac{2}{3}
\sin{\theta_W} - \frac{1}{2})
\end{eqnarray*}
\underline{\large{3. One-loop corrections to the parameter $\mu$} \cite{14}}
\begin{eqnarray*}
\Delta_Z &=& \frac{3g_2^2}{32 \pi^2} \frac{m_t^2}{M_W^2 \cos^2{\beta}}
[ b(\tilde{m}_{t_1}^2) + b(\tilde{m}_{t_2}^2) + 2 m_t^2 + \\
 &+& (A_t^2- \mu^2 \cot^2{\beta}) \frac{b(\tilde{m}_{t_1}^2) -
b(\tilde{m}_{t_2}^2)}{\tilde{m}_{t_1}^2 - \tilde{m}_{t_2}^2} ]
\end{eqnarray*}
where $b(m)= m^2 (\log\frac{m^2}{m_t^2}-1)$

\vglue 0.5cm

{\bf Acknowledgments}

\vglue 0.5cm

We are grateful to W. de Boer, K.A.Ter-Martirosian and  R.B.Nevzorov
for useful discussions.  Financial support from RFBR grant \#
96-02-17379 is kindly acknowledged. Also this work was partially
supported by INTAS under the Contract INTAS-96-155. One of the authors (G.
Y.) is thankful to the Joint Institute for Nuclear Research (Dubna)
where this work has been started, for a hospitality.

\end{document}